\documentclass[twocolumn,doublespacing,showpacs,preprintnumbers,amsmath,amssymb]{revtex4}
\usepackage{}
\usepackage{amssymb}
\usepackage{txfonts}
\usepackage{graphicx}
\usepackage{subfigure}
\usepackage{dcolumn}
\usepackage{bm}
\usepackage{sidecap}

\begin{document}
\title{Improve the Maximum Transmission Distance of Four-State Continuous Variable Quantum Key Distribution by using a Noiseless Linear Amplifier}
\author{Bingjie Xu$^{1,2}$}
\thanks{xbjpku@pku.edu.cn}
 \author{Chunming Tang$^{2}$}
 \author{Hui Chen$^{1,2}$}
 \author{Wenzheng Zhang$^{1,2}$}
 \author{Fuchen Zhu$^{1,2}$}
 \affiliation{1. Science and Technology on Security Communication Laboratory, 610041, Cheng Du, China\\
2. Institute of Southwestern Communication, 610041, Cheng Du, China}
\date{\today}
\begin{abstract}
A modified four-state CVQKD protocol is proposed to increase the maximum transmission distance and tolerable excess noise in the presence of Gaussian lossy and noisy channel by using a noiseless linear amplifier (NLA). We show that a NLA with gain $g$ can increase the maximum admission losses by $20\log_{10}g$ dB.
\end{abstract}
\maketitle

\section{Introduction}
Quantum key distribution (QKD) provides a means of sharing a secret key between two parties (Alice and Bob) securely in the presence of an eavesdropper (Eve) ~\cite{RMP_09,{Rev_Lo_08}}. The single-photon (e. g. BB84~\cite{BB84}), entanglement-based (e. g. E91~\cite{E91}) and continuous variable (e. g. GG02 \cite{GG02}) QKD protocols have proved to be unconditionally secure under some (e. g. source, detection, and post-processing) assumptions \cite{RMP_09}. In the continuous-variable QKD (CVQKD) protocols, information is encoded in quadratures of coherent or squeezed states, and decoded by homodyne or heterodyne detections~\cite{GG02,NoSwitching,SquzeedProtocol,SquzeedProtocol2}, which have the advantage of only requiring off-the-shelf telecom components~\cite{RMP_12}. Besides experimental demonstrations~\cite{CV_exp,{CV_exp2}}, the theoretical security of CVQKD has been established against general collective Gaussian attacks~\cite{CVQKD1,CVQKD2,CVQKD3,CVQKD4}, which has been shown optimal in the asymptotical limit~\cite{CVQKD5}. Furthermore, the effect of finite size has been recently investigated in CVQKD protocol~\cite{CV_FinitSize,CV_FinitSize2,CV_FinitSize3}.

Developing QKD protocol resistant to loss and noise is of great practical importance.  The CVQKD protocol based on Gaussian modulation of coherent state has been proven to be technically practical~\cite{RMP_12}. However, due to the low reconciliation efficiency for correlated Gaussian variables at low SNR (signal to noise ration), the maximum transmission distance of CVQKD is quite limited~\cite{CV_exp2}. There are two possible ways to solve this problem. One is to build good reconciliation algorithms with reasonable efficiency even at low SNR for Gaussian modulation protocols, where steady progress has been made in recent years~\cite{CV_Recon}. The other is to use discrete modulation CVQKD~\cite{DMCVQKD1,DMCVQKD2,DMCVQKD3,DMCVQKD4}, such as the four-state protocol~\cite{DMCVQKD2,DMCVQKD3}, which has been proved to be secure against collective attacks and have large enough reconciliation efficiency at low SNR.

Recently, it is very interesting to see that one can improve the maximum transmission distance of Gaussian modulation CVQKD protocols dramatically by
using a nondeterministic noiseless linear amplifier (NLA)~\cite{CV_NLA}. Lately, a method was proposed to improve the secret key rate of four-state CVQKD protocol over long distance by using a phase-sensitive or phase-insensitive optical amplifier, while the maximum transmission distance is decreased~\cite{CV_LA}. Inspired by the methods in~\cite{CV_NLA}, we show that the maximum transmission distance and tolerable excess noise of the four-state CVQKD protocol can be increased by using a NLA before Bob's detection in the presence of a lossy and noisy Gaussian channel. Similar to the result in~\cite{CV_NLA}, we find that a NLA with gain $g$ can increase the maximum admissible losses by a factor of $g^{-2}$. Because of the nondeterministic nature of the NLA, the security proof here is similar to that in CVQKD protocols with post-selection.

\section{The four-state CVQKD  Protocol}
In this section, we firstly describe the prepare-and-measure (PM) and entanglement-based (EB) version of the four-state CVQKD protocol. Then, the secure key rate for the protocol under collective attack is given in detail.
\subsection{The PM and EB description of four-state CVQKD protocol}
In the PM version of the four-state CVQKD protocol~\cite{DMCVQKD2}, Alice sends randomly one of the four coherent states $\left\{\left| {\alpha_k}\right\rangle=\left| {\alpha {e^{i\pi (2k+1)/4}}}\right\rangle, k=0,1,2,3\right\}$ with probability $1/4$ to Bob through a quantum channel, where $\alpha$ is chosen to be a real positive number. Bob measures randomly one of the quadratures in homodyne detection, and decodes the information by the sign of his measurement result.

The PM version of the four-state CVQKD protocol can be reformulated in EB version. Alice initially prepare a two-mode entangled state
\begin{equation}
\left| \Phi_{AB}(\alpha)  \right\rangle  = \frac{1}{2}\sum\limits_{k = 0}^3 {\left| {{\psi _k}} \right\rangle_{A} \left| {\alpha {e^{i(2k + 1)\pi /4}}} \right\rangle_{B} },
\end{equation}
where \begin{equation}\left| {{\psi _k}} \right\rangle  = \frac{1}{2}\sum\limits_{m = 0}^3 {{e^{i(1 + 2k)m\pi /4}}\left| {{\phi _m}} \right\rangle }\ (k=0,1,2,3)\end{equation}is a non-Gaussian orthogonal state, and
 \begin{eqnarray*}
 \left| {{\phi _k}} \right\rangle  &=& \frac{{{e^{ - {\alpha ^2}/2}}}}{{\sqrt {{\lambda _k}} }}\sum\limits_{n = 0}^\infty  {{{( - 1)}^n}\frac{{{\alpha ^{4n + k}}}}{{\sqrt {(4n + k)!} }}} \left| {4n + k} \right\rangle , \\
 {\lambda _{0,2}} &=& \frac{1}{2}{e^{ - {\alpha ^2}}}[\cosh ({\alpha ^2}) \pm \cos ({\alpha ^2})],\\
 {\lambda _{1,3}} &=& \frac{1}{2}{e^{ - {\alpha ^2}}}[\sinh ({\alpha ^2}) \pm \sin ({\alpha ^2})].
\end{eqnarray*}
Then Alice performs a projective measurement $\{ \left| {{\psi _0}} \right\rangle \left\langle {{\psi _0}} \right|$, $\left| {{\psi _1}} \right\rangle \left\langle {{\psi _1}} \right|$, $\left| {{\psi _2}} \right\rangle \left\langle {{\psi _2}} \right|$, $\left| {{\psi _3}} \right\rangle \left\langle {{\psi _3}} \right|\}$ on mode A to project mode B on one of the four coherent states $\left| {\alpha_k} \right\rangle (k=0,1,2,3)$ randomly, which are then measured by a homodyne detector at Bob's side after passing through a quantum channel.

Although the EB version does not correspond to the actual implementation, it is fully equivalent to the PM version from the a secure point of view~\cite{DMCVQKD2,DMCVQKD3}, and it provides a powerful description of establishing security proof against collective attacks through the covariance matrix $\gamma_{AB}$ of the state before their respective measurements~\cite{CV_LA,CV_NLA}.

\subsection{Secure key rate of four-state CVQKD protocol}
The covariance matrix $\gamma_{A_0B_0}$ of the state $\left| \Phi_{AB} (\alpha) \right\rangle$ is
\begin{equation}
\gamma_{A_0B_0}=\left[\begin{array}{*{20}{c}}
   V\mathbb{I} & Z\sigma_z
  \\
  Z\sigma_z & V\mathbb{I}  \\
\end{array}\right],
\end{equation}
where $\mathbb{I}=\left[\begin{array}{*{20}{c}}
   1 & 0\\
   0 & 1\\
\end{array}\right]$ and $\sigma_z=\left[\begin{array}{*{20}{c}}
   1 & 0\\
   0 & -1\\
\end{array}\right]$, $V=2\alpha^2+1=V_A+1$ is variance of quadratures for mode A and B, and
\begin{equation}
Z=2\alpha^2(\lambda_0^{3/2}\lambda_1^{-1/2}+\lambda_1^{3/2}\lambda_2^{-1/2}+\lambda_2^{3/2}\lambda_3^{-1/2}+\lambda_3^{3/2}\lambda_0^{-1/2})
\end{equation}
reflects the correlation between mode A and mode B. After mode B passing through a Gaussian channel with transmittance $T$ and equivalent excess noise at the input $\epsilon$, the quantum state $\left| \Phi_{AB} (\alpha) \right\rangle$ turns to state $\rho_{AB}$ with covariance matrix
\begin{equation}
\gamma_{AB}(\alpha,T,\epsilon)=\left[\begin{array}{*{20}{c}}
   V\mathbb{I} & \sqrt{T}Z\sigma_z\\
  \sqrt{T}Z\sigma_z & T(V+\chi)\mathbb{I}  \\
\end{array}\right],
\end{equation}
where $\chi=\frac{T}{1-T}+\epsilon$ is the equivalent total noise at the input. This matrix contains all the information needed to establish the secret key rate for collective attacks  for the four-state CVQKD protocol, and the lower bound of secure key rate with reverse reconciliation is~\cite{DMCVQKD2}
\begin{equation}
R(\alpha,T,\epsilon)\ge\beta I_{AB}(\alpha,T,\epsilon)-S_{BE}(\alpha, T, \epsilon),
\end{equation}
where
\begin{equation}
I_{AB}(\alpha,T,\epsilon)=\frac{1}{2}\log_2(\frac{V+\chi}{1+\chi})=\frac{1}{2}\log_2(1+{\rm SNR})
\end{equation}
refers here to the mutual information between Alice and Bob~\cite{CV_LA}, $S_{BE}$ is the Holevo bound for the mutual information shared by Eve and Bob, and $\beta<1$ is the reconciliation efficiency ($\beta>0.8$ can be reached for arbitrary low SNR in the four-state CVQKD protocol~\cite{DMCVQKD2,DMCVQKD3}). As shown in~\cite{DMCVQKD2}, $S_{BE}$ is maximized when the state $\rho_{AB}$ shared by Alice and Bob is a Gaussian state, which means $S_{BE}$ is upper bounded by the same quantity computed for a Gaussian state $\rho_{AB}^G$ with the same covariance matrix as the state $\rho_{AB}$ in an EB version of the protocol~\cite{DMCVQKD3}. Clearly, one has
\begin{equation}
{S_{BE}}\le{S_{BE}^G}= G\left( {\frac{{{v_1} - 1}}{2}} \right) + G\left( {\frac{{{v_2} - 1}}{2}} \right) - G\left( {\frac{{{v_3} - 1}}{2}} \right),
\end{equation}
where $G(x)=(x+1)\log_2(1+x)+x\log_2x$, and
\begin{equation}
{v_{1,2}} = \sqrt {\frac{1}{2}(\Delta \pm \sqrt {{\Delta^2} - 4D} )}
\end{equation}
are the symplectic eigenvalues of the covariance matrix $\gamma_{AB}$ where $\Delta=V^2+T^2(V+\chi)^2-2TZ^2$ and $D=(TV^2+TV\chi-TZ^2)^2$~\cite{DMCVQKD3,CV_LA},
and
\begin{equation}
{v_3} = \sqrt {V({V_A} + 1 - \frac{{T{Z^2}}}{{T{V_A} + 1 + T\epsilon }})}.
\end{equation}
is the symplectic eigenvalues of $\gamma_{A|B}$ which corresponds to the covariance matrix of Alice's state given the result of  Bob's homodyne measurement~\cite{DMCVQKD3}. Finally, the lower bound of secure key rate is
\begin{equation}\label{eq:origialkeyrate}
\underline{R}(\alpha,T,\epsilon)=\beta I_{AB}(\alpha,T,\epsilon)-S_{BE}^G(\alpha, T, \epsilon)\le R(\alpha,T,\epsilon).
\end{equation}

\section{Modified four-state CVQKD protocol by using a NLA}
\begin{figure}[b]
\begin{center}
\includegraphics[width=0.5\textwidth]{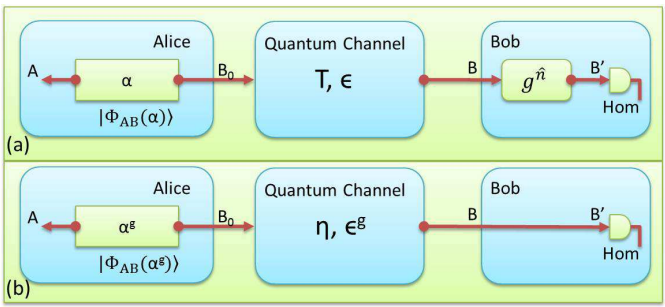}
\end{center}
\caption{(color online) (a) Modified four-state CVQKD protocol and (b) virtually equivalent protocol in EB version. A state $\left|\Phi_{AB}(\alpha)\right\rangle$ sent through a Gaussian channel with transmittance $T$ and excess noise $\epsilon$, followed by a successful amplification, has the same covariance matrix $\gamma_{AB}$ than a state $\left|\Phi_{AB}(\alpha^g)\right\rangle$ sent through a Gaussian with transmittance $\eta$ and excess noise $\epsilon^g$ without the NLA. The lower bound of secret key rate corresponding to successful amplified events in the modified four-state protocol $\underline{R^g}(\alpha,T,\epsilon)$ is the same as that of the virtually equivalent protocol $\underline{R}(\alpha^g, \eta, \epsilon^g)$.}
\label{fig:new scheme}
\end{figure}

Inspired by the method in~\cite{CV_NLA}, we propose a modified four-state CVQKD protocol by using a NLA as shown in Fig.~\ref{fig:new scheme}(a), where Alice and Bob implement the original four-state CVQKD protocol as usual but Bob adds a NLA before his homodyne detection, which is here assumed to be perfect for simplify of analysis as in~\cite{CV_NLA}.

A NLA can in principle probabilistic amplify the amplitude of a coherent state while retaining the initial level of noise~\cite{NLA1,NLA2,NLA3,NLA4,NLA5}. The successful amplification can be described by an operator $\hat{C}=g^{\hat{n}}$, where $\hat{n}$ is the photon number operator. When a NLA succeeds amplifying a coherent state,
\begin{equation}
\hat{C}\left| \alpha  \right\rangle=e^{ \frac{{{{\left| \alpha  \right|}^2}}}{2}({g^2} - 1) }\left| {g\alpha } \right\rangle,
\end{equation}
where $g$ is the amplitude gain of a NLA. In the modified four-state CVQKD protocol, only the events corresponding to a successful amplification will be used to exact a secret key, while the other events are aborted. Since the secure key rate of the protocol depends only on the covariance matrix $\gamma_{AB}$, it is sufficient to compute it in presence of the NLA to estimate the lower bound of secure key rate $\underline{R^g} (\alpha,T, \epsilon,g)$ corresponding to successfully amplified events.

In the following, we will analysis the performance of the modified four-state CVQKD protocol in two cases: (I) a lossy Gaussian channel without excess, where one can see directly the effect of the NLA; (II) a lossy and noisy Gaussian channel, which is a general and practical case.

\subsection{Case I: Lossy Gaussian channel without excess noise ($\epsilon=0$)}
After mode B passing through a lossy Gaussian channel with transmittance $T$ and no excess noise ($\epsilon=0$), the state $\left|\Phi_{AB}\right\rangle$ turns to $\left| {\Phi {'_{AB}}} \right\rangle  = \frac{1}{2}\sum\nolimits_{k = 0}^3 {\left| {{\psi _k}} \right\rangle \left| {\sqrt T {\alpha _k}} \right\rangle }$ with covariance matrix
\begin{equation}
\gamma_{AB}(\alpha,T,\epsilon=0)=\left[\begin{array}{*{20}{c}}
   V\mathbb{I} & \sqrt{T}Z\sigma_z\\
  \sqrt{T}Z\sigma_z & T(V+\frac{1-T}{T})\mathbb{I}  \\
\end{array}\right].
\end{equation}
Then a NLA with gain $g$ is added to amplify the input state at Bob's side on mode B, which can be described by the operator $g^{\hat{n}}$ when the input state is
successfully amplified. One can easily derive that 
\begin{equation}
g^{\hat{n}}\left| {\Phi {'_{AB}}} \right\rangle  = \frac{1}{2}e^{\frac{T}{2}\left|\alpha\right|^2(g^2-1)}\sum\limits_{k = 0}^3 {\left| {{\psi _k}} \right\rangle \left| {g\sqrt T {\alpha _k}} \right\rangle }.
\end{equation}
After the normalization of the output state, the successfully amplified quantum state is
$\left| {\Phi {''_{AB}}} \right\rangle  = \frac{1}{2}\sum\nolimits_{k = 0}^3 {\left| {{\psi _k}} \right\rangle \left| {g\sqrt T {\alpha _k}} \right\rangle }$ with covariance matrix
\begin{equation}
\gamma_{AB}^g(\alpha,T,\epsilon=0)=\left[\begin{array}{*{20}{c}}
   V\mathbb{I} & \sqrt{g^2T}Z\sigma_z\\
  \sqrt{g^2T}Z\sigma_z & g^2T(V+\frac{1-g^2T}{g^2T})\mathbb{I}  \\
\end{array}\right].
\end{equation}
One can find that the covariance matrix $\gamma_{AB}^g(\alpha,T,\epsilon=0)$ corresponds to successful amplification is equal to the covariance matrix $\gamma_{AB}(\alpha^g=\alpha,\eta=g^2T,\epsilon^g=0)$ of an equivalent system with $\left| \Phi_{AB}(\alpha)  \right\rangle$ sent through a channel with transmittance $\eta=g^2T$ and excess noise $\epsilon^g=0$, without using a NLA (as shown in Fig. 1(b)). Since the lower bound of secure key rate under collective attacks is completely determined by the covariance matrix shared by Alice and Bob, the lower bound of  secret key rate $\underline{R^g}(\alpha,T,\epsilon=0)$ corresponding to the successful amplified events is
\begin{equation}
\underline{R^g}(\alpha,T,\epsilon=0)=\underline{R}(\alpha,g^2T,\epsilon=0),
\end{equation}
and the lower bound of total secure key rate is
\begin{equation}
\underline{R^g_{tot}}(\alpha,T,\epsilon)=P_{success}\underline{R^g}(\alpha,T,\epsilon),
\end{equation}
where $P_{success}$ is the probability of successful amplification by a NLA. A direct conclusion is that a NLA with amplitude gain $g$ can increase the maximum admissible losses of the four-state CVQKD protocol by a factor $g^{-2}$ without excess noise, which is equivalent to improve the maximum transmission distance by $\frac{20\log_{10}g}{a}$ km, where $a=0.2dB/km$ for the fiber channel.

\subsection{Case II: Lossy and noisy Gaussian channel ($\epsilon > 0$)}
After passing through a Gaussian channel with transmittance $T$ and excess noise $\epsilon$, the state $\left|\Phi_{AB}(\alpha)\right\rangle$ turns to $\rho_{AB}$ with covariance matrix
\begin{equation}
\gamma_{AB}(\alpha,T,\epsilon)=\left[\begin{array}{*{20}{c}}
   V\mathbb{I} & \sqrt{T}Z\sigma_z\\
  \sqrt{T}Z\sigma_z &  T(V+\frac{1-T}{T}+\epsilon)\mathbb{I}  \\
\end{array}\right],
\end{equation}
where the quantum state of mode B is $\rho_B=\frac{1}{4}(\rho_0+\rho_1+\rho_2+\rho_3)$ with variance $T(V+\frac{1-T}{T}+\epsilon)$.

The $\rho_k$ is the state received by Bob if he knows Alice's measurement result on mode A is $k$, which can be described by a displaced thermal state
\begin{equation}
{\rho _k} = D({\beta_k}){\rho _{th}}(\lambda )D( - {\beta_k})=D(\sqrt T {\alpha_k}){\rho _{th}}(\lambda )D( - \sqrt T {\alpha_k}),
\end{equation}
where ${\rho _{th}}(\lambda ) = (1 - {\lambda ^2})\sum\nolimits_{n = 0}^\infty  {{\lambda ^{2n}}\left| n \right\rangle \left\langle n \right|}$ is a thermal state with variance
\begin{equation}
V(\lambda)=\frac{1+\lambda^2}{1-\lambda^2}=1+T\epsilon
\end{equation}
corresponds to Bob's variance when $V_A=0$. Then a NLA with gain $g$ is added to amplify the input state at Bob's side on mode B. Following the methods in~\cite{CV_NLA}, one can derive the effect of a NLA on the displaced thermal state $\rho_k (k=0,1,2,3)$. The P-function of a thermal state with parameter $\lambda$ is
\begin{eqnarray}
{\rho _{th}}(\lambda )
= \int {\frac{{1 - {\lambda ^2}}}{{{\pi\lambda ^2}}}{e^{ - \frac{{1 - {\lambda ^2}}}{{{\lambda ^2}}}{{\left| \alpha'  \right|}^2}}}} \left| \alpha'  \right\rangle \left\langle \alpha'  \right|d\alpha'.
\end{eqnarray}A displacement operation $D(\beta)$ will turn a thermal state to
\begin{equation}
\rho(\beta)=D(\beta){\rho _{th}}(\lambda )D(-\beta)=\int {\frac{{1 - {\lambda ^2}}}{{{\pi\lambda ^2}}}{e^{ - \frac{{1 - {\lambda ^2}}}{{{\lambda ^2}}}{{\left| \alpha'  -\beta\right|}^2}}}} \left| \alpha'  \right\rangle \left\langle \alpha'  \right|d\alpha'
\end{equation}
The effect of a NLA on displaced thermal state $\rho(\beta)$ is
\begin{eqnarray}
 \nonumber {\rho ^g}(\beta) &=& \hat C\rho(\beta) \hat C = \int {\frac{{1 - {\lambda ^2}}}{{\pi {\lambda ^2}}}{e^{ - \frac{{1 - {\lambda ^2}}}{{{\lambda ^2}}}{{\left| {\alpha'  - \beta } \right|}^2}}}} {g^{\hat n}}\left| \alpha'  \right\rangle \left\langle \alpha'  \right|{g^{\hat n}}d\alpha'  \\
  &=& \int {\frac{{1 - {\lambda ^2}}}{{\pi {\lambda ^2}}}{e^{ - \frac{{1 - {\lambda ^2}}}{{{\lambda ^2}}}{{\left| {\alpha'  - \beta } \right|}^2} + {{\left| \alpha'  \right|}^2}({g^2} - 1)}}} \left| {g\alpha' } \right\rangle \left\langle {g\alpha' } \right|d\alpha'.
\end{eqnarray}
By introducing $u=g\alpha'$, one gets
\begin{eqnarray}\label{eq:rhog}
 \nonumber {\rho ^g}(\beta)&=&\int {\frac{{1 - {\lambda ^2}}}{{g^2\pi {\lambda ^2}}}{e^{ - \frac{{1 - {\lambda ^2}}}{{{\lambda ^2}}}{{\left| {\frac{u}{g}  - \beta } \right|}^2+\left|u\right|^2\frac{g^2-1}{g^2}}}}} \left| {u } \right\rangle \left\langle {u } \right|du\\
 &=&C\int {{e^{ - \frac{{1 - {g^2\lambda ^2}}}{{{g^2\lambda ^2}}}{{\left| {u  - g\frac{1-\lambda^2}{1-g^2\lambda^2}\beta } \right|}^2}}}} \left| {u } \right\rangle \left\langle {u } \right|du,
\end{eqnarray}
where $C$ is a global unimportant normalization factor independent of the integrated variable $u$. The Eq.~(\ref{eq:rhog}) clearly correspond to a thermal state $\rho_{th}(g\lambda)$ displaced by $g\frac{1-\lambda^2}{1-g^2\lambda^2}\beta$. Thus, the successful amplification of $\rho_k$ corresponds to a new displaced thermal state ${\rho ^g_k}$,
\begin{equation}
{\rho ^g_k} = D(g\frac{{1 - {\lambda ^2}}}{{1 - {g^2}{\lambda ^2}}}{\beta _k}){\rho _{th}}(g\lambda )D( - g\frac{{1 - {\lambda ^2}}}{{1 - {g^2}{\lambda ^2}}}{\beta _k}),
\end{equation}
where $\beta_k=\sqrt{T}\alpha_k$. Finally, the action of the NLA on the input state at Bob's side introduce the transformations
\begin{eqnarray}
 \sqrt T {\alpha _k} \to g\frac{{1 - {\lambda ^2}}}{{1 - {g^2}{\lambda ^2}}}\sqrt T {\alpha _k},{\lambda ^2} \to {g^2}{\lambda ^2},
\end{eqnarray}
which is equivalent to
\begin{equation}\label{eq:trans1}
 \sqrt T {\alpha _k} \to\frac{2 g}{{2 - ({g^2} - 1)T\epsilon }}\sqrt T {\alpha _k},\frac{{T\epsilon }}{{2 + T\epsilon }} \to {g^2}\frac{{T\epsilon }}{{2 + T\epsilon }}.
\end{equation}

In the next step, one need to consider the action of NLA when Bob does not know Alice's measurement result.  In such a case, the input state of the NLA is
\begin{equation}
\rho_B=\frac{1}{4}(\rho_0+\rho_1+\rho_2+\rho_3),
\end{equation}
with variance $1+T\epsilon+TV_A=\frac{1+\lambda^2}{1-\lambda^2}+2T\alpha^2$. The output state corresponding to successful amplified events on $\rho_B$ is
\begin{equation}
{\rho _{B'}} = \frac{1}{4}\left( {{\rho ^g_0} + {\rho ^g_1} + {\rho ^g_2} + {\rho ^g_3}} \right),
\end{equation}
with variance $\frac{1+g^2\lambda^2}{1-g^2\lambda^2}+2g^2(\frac{1-\lambda^2}{1-g^2\lambda^2})^2T\alpha^2$. Thus, the  action of the NLA on the input state at Bob's side introduce the transformations
\begin{equation}\label{eq:trans2}
\frac{{1 + {\lambda ^2}}}{{1 - {\lambda ^2}}} + 2T{\alpha ^2} \to \frac{{1 + {g^2}{\lambda ^2}}}{{1 - {g^2}{\lambda ^2}}} + 2{g^2}{\left( {\frac{{1 - {\lambda ^2}}}{{1 - {g^2}{\lambda ^2}}}} \right)^2}T{\alpha ^2},
\end{equation}
which can be derived from Eq.~(\ref{eq:trans1}).

\begin{figure}[t]
\begin{center}
\includegraphics[width=0.49\textwidth]{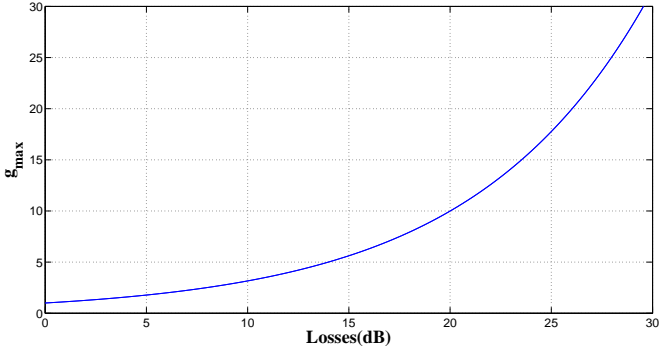}
\end{center}
\caption{(color online) The maximum value of the gain of NLA $g_{max}$ against the losses when $\epsilon=0.02$. }
\label{fig:gmax}
\end{figure}

Now we want to find a virtual two-mode entangled state $\left|\Phi_{AB}(\alpha^g)\right\rangle$, sent through a channel of transmittance $\eta$ and excess noise $\epsilon^g$ without using the NLA, while share the same covariance matrix as $\gamma^g_{AB}(\alpha,T,\epsilon)$ as shown in Fig. 1(b). The following conditions should be satisfied
 \begin{eqnarray}
 \label{eq:condition1} \sqrt \eta  {\alpha^g} &=& g\frac{2}{{2 - ({g^2} - 1)T\epsilon }}\sqrt T {\alpha} , \\
 \label{eq:condition2} \frac{{\eta {\epsilon ^g}}}{{2 + \eta {\epsilon ^g}}} &=& {g^2}\frac{{T\epsilon }}{{2 + T\epsilon }}, \\
 \label{eq:condition3} 1 + \eta {\epsilon ^g} + 2\eta{{\alpha ^g}^2} &=& \frac{{1 + {g^2}{\lambda ^2}}}{{1 - {g^2}{\lambda ^2}}} + 2{g^2}{\left( {\frac{{1 - {\lambda ^2}}}{{1 - {g^2}{\lambda ^2}}}} \right)^2}T{\alpha ^2},
 \end{eqnarray}
 where the third line can be derived by the first two lines. So, one only need to consider the conditions in Eqs.~(\ref{eq:condition1}) and~(\ref{eq:condition2}). Remind that one always has $\alpha^g=\alpha$ for $\epsilon=0$ or $g=1$, so we add a third condition
 \begin{equation} \label{eq:condition4}
 \alpha^g=\alpha.
 \end{equation}
 Then the solutions for Eqs. ~(\ref{eq:condition1}),~(\ref{eq:condition2}), and~(\ref{eq:condition4}) is
 \begin{eqnarray}
 \eta = \frac{4g^2T}{[2 + (1 - {g^2})T\epsilon]^2},{\epsilon ^g} = \epsilon  - \frac{1}{2}({g^2} - 1)T\epsilon^2,\alpha^g=\alpha.
 \end{eqnarray}
Those parameters can be interpreted as physical parameters of an equivalent system if they satisfy the physical meaning constrains $0\le\eta\le1$ and $\epsilon^g\ge0$, which require
 \begin{eqnarray*}
{g_{\max }}(T,\epsilon ) = \left\{ {\begin{array}{*{20}{c}}
   {\frac{1}{{\sqrt T }},\epsilon  = 0}  \\
   {\frac{{ - 2\sqrt T  + \sqrt {4T + 4T\epsilon (2 + T\epsilon )} }}{{2T\epsilon }},\epsilon  > 0}  \\
\end{array}} \right.
 \end{eqnarray*}
which is plotted in Fig.~\ref{fig:gmax}. Finally, one has
\begin{eqnarray}
\underline{R^g}(\alpha,T,\epsilon)&=&\underline{R}(\alpha,\eta,\epsilon^g),\\ \underline{R^g_{tot}}(\alpha,T,\epsilon)&=&P_{success}\underline{R^g}(\alpha,T,\epsilon).\label{eq:modifiedkeyrate}
\end{eqnarray}

\section{Numerical Simulation and Discussion}
In the following, the performance of the modified four-state protocol is compared with the original one for a given channel with the same transmittance $T$ and excess noise $\epsilon$. The secure key rate of the original protocol is given by $\underline{R}(\alpha,T,\epsilon)$ in Eq.~(\ref{eq:origialkeyrate}), and that of modified protocol is given by $\underline{R^g_{tot}}(\alpha,T,\epsilon)$ in Eq.~(\ref{eq:modifiedkeyrate}).
In numerical simulations, $P_{success}$ is assumed to be a constant, which is reasonable when $\beta<1$~\cite{CV_NLA}. The precise value of $P_{success}$ depends on practical implementations. However, it is not not important for our result, since it only acts as a scaling factor of $\underline{R^g_{tot}}(\alpha,T,\epsilon)$ and does not change the fact that a negative secret rate $\underline{R}(\alpha,T,\epsilon)$ can become positive $\underline{R^g_{tot}}(\alpha,T,\epsilon)$ with a NLA for a certain distance of transmission distance. As shown in~\cite{CV_NLA}, the value of $P_{success}$ is upper bounded by $1/g^2$, and we choose the value $P_{success}=1/g^2$ to optimize the performance of the modified four-state CVQKD protocol.

\begin{figure*}[t]
\begin{center}
\includegraphics[width=0.90\textwidth]{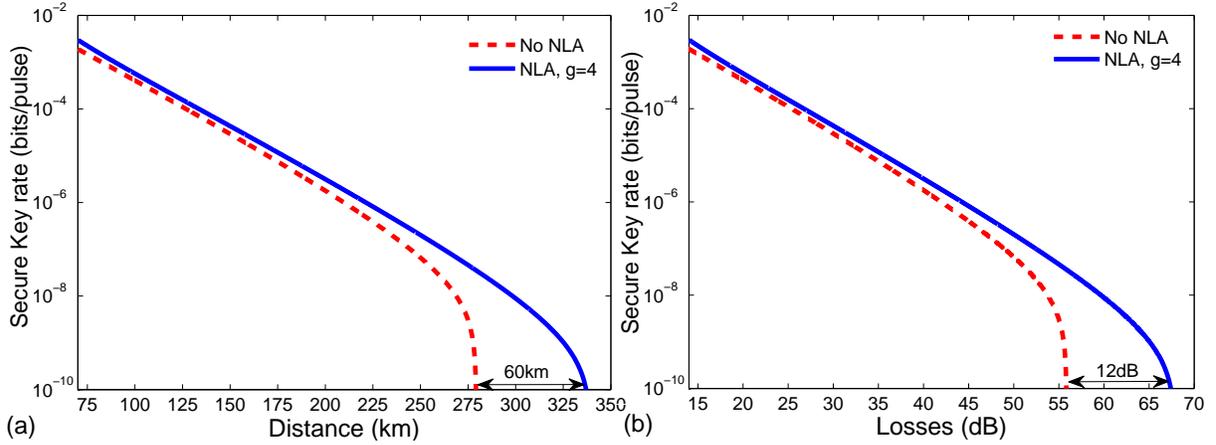}
\end{center}
\caption{(color online) (a). The lower bound of secret key rate for the modified four-state CVQKD protocol with a NLA $\underline{R^g_{tot}}(\alpha,T,\epsilon)$ and that of the original protocol without a NLA $\underline{R}(\alpha,T,\epsilon)$ against the transmission distance in $km$. (b). The lower bound of secret key rate for the modified four-state CVQKD protocol with a NLA $\underline{R^g_{tot}}(\alpha,T,\epsilon)$ and that of the original protocol without a NLA $\underline{R}(\alpha,T,\epsilon)$ against the losses in $dB$. In the simulations, $V_A=2\alpha^2=0.25$, $\epsilon=0.002$, $\beta=0.8$, $g=4$, and $P_{success}=1/g^2$.}
\label{fig:performance}
\end{figure*}

\begin{figure}[htbp]
\begin{center}
\includegraphics[width=0.48\textwidth]{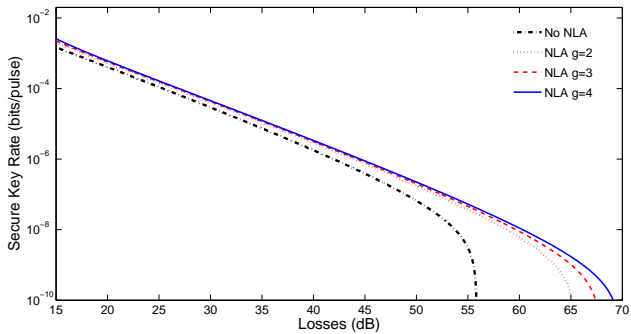}
\end{center}
\caption{(color online) The lower bound of the secret key rate for the modified four-state CVQKD protocol with a NLA $\underline{R^g_{tot}}(\alpha,T,\epsilon)$ for $g=2,3,4$ and that of the original protocol without a NLA $\underline{R}(\alpha,T,\epsilon)$ against the losses in $dB$. In the simulations, $V_A=2\alpha^2=0.25$, $\epsilon=0.002$, $\beta=0.8$, and $P_{success}=1/g^2$. }
\label{fig:optimal_g}
\end{figure}

The numerical simulations of $\underline{R^g_{tot}}(\alpha,T,\epsilon)$ and $\underline{R}(\alpha,T,\epsilon)$ with the same channel parameters $T$ and $\epsilon$ is shown in Fig.~\ref{fig:performance}, where the amplitude gain of the NLA is $g=4$ and the excess noise is $\epsilon=0.002$. Similar to the results in~\cite{CV_NLA}, one can find that the maximum admissible losses is increased by $20\log_{10}g$ by using a NLA with gain $g$,  which is equivalent to increase the maximum transmission distance by $\frac{20\log_{10}g}{0.2}$ km in fiber channel. This result does not depends on the value of $P_{success}$. Even for a more realistic probability of success, the NLA increase the maximum transmission distance in the same way. To test the efficiency for different values of $g$, the secret key rate for the modified four-state CVQKD protocol with a NLA with gain $g=2,3,4$ is compared with that of the original protocol, as shown in Fig.~\ref{fig:optimal_g}.

\begin{figure}[htbp]
\begin{center}
\includegraphics[width=0.48\textwidth]{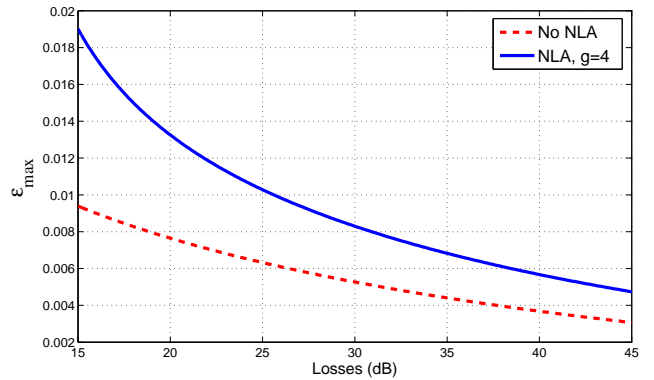}
\end{center}
\caption{(color online) Maximal excess noise for the modified and original four-state CVQKD protocol against losses in $dB$. In the simulations, $V_A=2\alpha^2=0.25$, $\beta=0.8$, $g=4$, and $P_{success}=1/g^2$.}
\label{fig:excessnoise}
\end{figure}

The maximal tolerable excess noise $\epsilon_{\max}$ for the modified four-state CVQKD protocol by using a NLA with gain $g$ and that of the original protocol is shown in Fig.~\ref{fig:excessnoise}. By using a NLA, the maximal tolerable excess noise can be increased, and this result does not depend on the value of $P_{success}$.

\section{Conclusion}
In this paper, a modified four-state CVQKD protocol is proposed. The maximum transmission distance can be increased by the equivalent of $20\log_{10} g$ dB of losses by using a noiseless linear amplifier before Bob's detection. The modified protocol is also more robust against excess noise.

Steady progress on the experimental realization of the NLA has been made in recent years~\cite{NLA1,NLA2,NLA3,NLA4,NLA5}. A further work would be analyzing  the gaps between practical implementations and theoretical description of NLA, and the effect of the imperfection on the secure key rate.

This work is supported by Foundation of Science and Technology on Communication Security Laboratory (Grants No. 9140c11010110c1104).

\end{document}